\documentclass{article}

\usepackage{arxiv}

\usepackage[utf8]{inputenc} 
\usepackage[T1]{fontenc}    
\usepackage{hyperref}       
\usepackage{url}            
\usepackage{booktabs}       
\usepackage{amsfonts}       
\usepackage{nicefrac}       
\usepackage{microtype}      
\usepackage{lipsum}		

\usepackage{graphicx}
\usepackage{amsmath}
\usepackage{multirow}
\usepackage[table]{xcolor}
\usepackage{array}
\usepackage{slashbox}
\usepackage{pict2e}
\usepackage{float}
\newcolumntype{R}[1]{>{\raggedleft\arraybackslash }b{#1}}
\newcolumntype{L}[1]{>{\raggedright\arraybackslash }b{#1}}
\newcolumntype{C}[1]{>{\centering\arraybackslash }b{#1}}

\title{Vine disease detection in UAV multispectral images with deep learning segmentation approach}

\author{
  Mohamed Kerkech \\
  INSA-CVL, Univ. Orl\'{e}ans\\ 
  PRISME, EA 4229\\
  F18022, Bourges, France\\
  \texttt{mohamed.kerkech@insa-cvl.fr} \\
   \And
  Adel Hafiane\\
  INSA-CVL, Univ. Orl\'{e}ans\\ 
  PRISME, EA 4229\\
  F18022, Bourges, France\\
  \texttt{adel.hafiane@insa-cvl.fr} \\
	\And   
  Raphael Canals \\
  Univ. Orl\'{e}ans, INSA-CVL\\
  PRISME, EA 4229\\
  F45072, Orl\'{e}ans, France \\
  \texttt{raphael.canals@univ-orleans.fr} \\
}

\begin{document}
\maketitle

\begin{abstract}
One of the major goals of tomorrow's agriculture is to increase agricultural productivity but above all the quality of production while significantly reducing the use of inputs. Meeting this goal is a real scientific and technological challenge. Smart farming is among the promising approaches that can lead to interesting solutions for vineyard management and reduce the environmental impact. Automatic vine disease detection can increase efficiency and flexibility in managing vineyard crops, while reducing the chemical inputs. This is needed today more than ever, as the use of pesticides is coming under increasing scrutiny and control. The goal is to map diseased areas in the vineyard for fast and precise treatment, thus guaranteeing the maintenance of a healthy state of the vine which is very important for yield management. To tackle this problem, a method is proposed here for vine disease detection using a deep learning segmentation approach on Unmanned Aerial Vehicle (UAV) images. The method is based on the combination of the visible and infrared images obtained from two different sensors. A new image registration method was developed to align visible and infrared images, enabling fusion of the information from the two sensors. A fully convolutional neural network approach uses this information to classify each pixel according to different instances, namely, shadow, ground, healthy and symptom. The proposed method achieved more than $92\%$ of detection at grapevine-level and $ 87\% $ at leaf level, showing promising perspectives for computer aided disease detection in vineyards.
\end{abstract}

\keywords{Unmanned aerial vehicle (UAV) \and image registration \and convolutional neural network \and precision agriculture \and disease mapping.}

\section{Introduction}
Several studies have been carried out on the overuse of crop pesticides and their negative effects on human health~\cite{Aktar2009,Pimentel1993,Margalida2014}. Like other crops, vines are very vulnerable to viruses, bacteria and fungi. This vulnerability favours their contamination by several types of disease that are harmful and destructive~\cite{Tyerman2009}, such as \textit{Esca}~\cite{Hofstetter2012}, \textit{Flavescence dorée}~\cite{Chuche2014} and Mildew~\cite{Gessler2011}. Vine contamination generally reduces productivity~\cite{macdonald_remote_2016-1}, which implies economic losses for the winegrower. To deal with this situation, winegrowers have to frequently check the state of the vine leaves. However, this traditional procedure is laborious and expensive, since it involves several experts for many days~\cite{Brousse2016}. To reduce economic loss and the environmental impact, remote sensing methods are a promising approach for effective vineyard monitoring.

Remote sensing of agricultural crops~\cite{Teke2013} has evolved considerably over the past decade. Applications such as calculating fertilizer rates~\cite{Schut2018}, monitoring biomass production~\cite{Karpina2016}, weed detection~\cite{Bah2017}, detecting defective crops~\cite{Serif2019} or disease detection~\cite{Khirade2015,Pinto2017,Schor2016} have been proposed. These applications are constantly progressing as technology advances, especially with the evolution of Unmanned Aerial Vehicles (UAV) which have opened up further research opportunities thanks to their low manufacturing costs.

UAVs are increasingly used in many fields, such as urban remote sensing, but also in a wide range of agricultural applications~\cite{Garcia2019}. Previous studies have shown the importance of both the visible and the infrared spectrum for disease detection~\cite{Chen2018}. Combining these two imaging modalities would therefore ensure better detection. In UAV imaging systems, usually two separate sensors are used, one for each modality. However, an acquisition by two sensors generates a spatial shift between the visible and infrared image which makes it difficult to process the information from the two sensors simultaneously. Therefore, multimodal alignment or registration~\cite{Penaranda2017,Wang2018,Tsai2017,Inostroza2017} is required to fuse both sensors information with deep learning.

Deep learning techniques have enabled great progress in the computer vision field thanks to the convolutional neural networks (CNNs) approach~\cite{ LeCun1998,Krizhevsky,Zeiler2014,Simonyan2014,Szegedy2014,He}. As in several fields of application, these technologies are increasingly used in the remote sensing domain for agriculture~\cite{Kamilaris2018,Applications2019,Sladojevic2016,Lu2017,Yang2019,Fuentes2017,Lee2019,Bah2018,Kerkech2018}. Most research on agricultural applications uses CNNs with the sliding window technique, which generally leads to fuzzy boundaries of the image regions. On the other hand, crop disease detection can be seen as an image segmentation problem. Therefore, one can benefit from the deep learning segmentation approach to detect crop disease with a better boundary precision compared to the sliding window technique. Several segmentation architectures have been developed, such as SegNet~\cite{Badrinarayanan}, DeconvNet~\cite{Noh} and U-Net~\cite{Ronneberger}. SegNet is the most popular architecture for semantic segmentation~\cite{Kendall,Murillo,Images2017,Guo2018a}. It has shown a very good performance in solving problems related to semantic segmentation for several applications~\cite{Nguyen2019,Du1998,Wei2019}. So far, very little attention has been paid to the role of deep learning segmentation for vine disease detection.

This paper presents a new methodology for vine disease detection in aerial images, using multispectral information. The aim was to develop algorithms and methods in order to investigate the possibility of detecting vine diseases, using the potential of deep learning segmentation architecture. The problem was addressed by the semantic segmentation approach in order to identify classes such as shadow, ground, healthy and symptomatic vines. The method consists in two main steps. The first one deals with the problem of multispectral image registration, where a new method was proposed to effectively align images from the visible and infrared spectres. The second one uses the SegNet architecture to delineate semantic areas in each image type separately, then a fusion procedure was applied to the segmentation outputs. Data were collected under real conditions on two vineyard plots. Several experimental schemes have been set up to show the contribution of different elements of the proposed method. The study provides an important insight into the potential of recent machine learning approaches for disease mapping using UAV remote sensing technology. 

The article is organized as follows: related work is presented in section~\ref{StateOfTheArt}, the study areas and materials are described in section~\ref{AreasAndMaterials}, the proposed methods are detailed in section~\ref{MaterialsAndMethods}, the experiments and results are presented in section~\ref{Experimentation}, the proposed system are discussed in section~\ref{Discussion} and section~\ref{Conclusion} concludes the paper.\\

\section{Related work}	
\label{StateOfTheArt}
This section summarizes main studies carried out on image registration, and disease detection in vineyards, plants or crops.

\subsection{Image registration}	
In the literature, the problem of image registration dates back to the 1980s. Since then, several methods have been implemented. The work accomplished in various fields has been surveyed in: medicine~\cite{Ferrante2017}, computer vision~\cite {Flusser2003}, remote sensing~\cite {Dawn2010} and various applications~\cite {Nag}. In all areas studied, it is concluded that image registration is based on two main methods: the area-based method, and the feature-based method.

\textit{The area-based method:} This method is not widely used in the remote sensing field, because most of the algorithms are highly sensitive to several uncontrollable parameters, such as variations in brightness, image noise, etc. The method is therefore generally applicable only to non-rigid problems. However, Wang et al.~\cite{Wang2010} implemented an algorithm (An automatic cross-correlation (ACC)) insensitive to the light conditions and applied it to multimodal images (visible and infrared). The authors concluded that this algorithm performs better than other area-based algorithms and is suitable for multimodal images. Another registration method applied to precision agriculture was proposed by Erives et al.~\cite {Erives2005}. This method that processes multispectral images is based on the phase correlation algorithm (PC). The results obtained indicate that this algorithm is robust to modality change, brightness difference, noise, rotation and translation. Zhuang et al.~\cite {Zhuang2016} performed a multimodal registration based on mutual information with a combination of Particle Swarm Optimization (PSO) and Powell search algorithms. The proposed method was found to be faster in terms of runtime and more accurate in terms of results compared to traditional methods.

\textit{The feature-based method:} Lakshmi et al.~\cite{Sreenu2015} and Javadi et al.~\cite{Javadi2015} worked on natural terrain and city video frames, acquired by a UAV for the creation of an orthorectified image. In these studies, the standard registration method based on the SURF algorithm was used. In~\cite{Sreenu2015}, this method was compared with the Cross-Correlation algorithm (Area-based method) which failed to register the aerial images, whereas the feature-based method using the variants of the SIFT algorithm outperformed the other methods in terms of results and in terms of runtime. Tsai et al.~\cite{Tsai2017} conducted a similar study to compare the SIFT algorithm with ABRISK, and concluded that the ABRISK algorithm was up to$312$times faster than SIFT, and had a lower mean error. In another study, Onyango et al.~\cite{Onyango2017} used the AKAZE algorithm to match oblique building images with images of cities taken by a UAV. The study concluded that the AKAZE algorithm outperformed other algorithms of the same type. More recently, image matching algorithms based on deep learning have appeared~\cite{Dan2018, Wang2018}. The deep learning architecture is used as a feature extractor to create a correspondence between the two images. Wang et al.~\cite{Wang2018} worked on remote sensing images using a supervised architecture, while Yang et al.~\cite{Dan2018} used an unsupervised architecture to recalibrate multi-temporal images. The latter showed better accuracy than the SIFT type algorithms, but these results only correspond to multi-temporal images, of the same modality and on low resolution images.

\subsection{Disease detection}	
A comprehensive review of the literature was conducted by Mahlein~\cite{mahlein_plant_2016}. The survey lists several studies on disease detection by multispectral imaging. Among others, Oerke et al.~\cite{Oerke2011} demonstrated that disease symptoms in the plants begin to appear in the infrared range several days before their appearance in the visible range. The potential of multispectral information in the early detection of plant disease is attracting more and more interest in the remote sensing field. Two studies on the hyperspectral reflectance of vine leaves diseased by complex \textit{Esca}, the first one~\cite{Palou2013} at the leaf level, and the second one~\cite{DiGennaro2016} at the vine field level (UAV images). Both showed a difference between a reflectance of a healthy and diseased leaf. 

In a first study, Albetis et al.~\cite{albetis_detection_2017} investigated the detection of \textit{Flavescence dorée} in UAV images. The study was carried out on plots of white and red cultivars. The results obtained indicate the feasibility of disease detection using aerial images. In a second study, Albetis et al.~\cite{Albetis2019} examined the potential of multispectral imaging by UAV for the detection of symptomatic and asymptomatic vines. In addition to the first study, a larger dataset was acquired and used to test $24$ variables calculated from this new dataset. The best results were obtained by the red-green index (RGI) and the red-green vegetation index (GRVI). 

In Al-Saddik et al.~\cite{Dor2017}, a first study on hyperspectral images at the vine leaf scale was carried out. The aim of the study was to develop spectral disease indices capable of detecting and identifying \textit{Flavescence dorée} disease in vines, and achieved $ 90\% $ classification accuracy. A second study by Al-Saddik et al.~\cite{Al-saddik2018} on disease identification at the vine leaf level was carried out to differentiate between yellowing leaves, and leaves infected with \textit{Esca} disease through a neural network classifier. The best results were obtained when textural and spectral data were combined. A third study by the same authors~\cite{Al-saddik2019} consisted in defining the most significant spectral channels for \textit{Flavescence dorée} disease detection.

Rançon et al.~\cite{Rancon2019} carried out a similar study on \textit{Esca} disease detection in vineyards. The imaging system was mounted on a small vehicle that passed between the vine rows for image acquisition. Two methods for detecting \textit{Esca} disease were used: Scale Invariant Feature Transform (SIFT) encoding, and the deep learning MobileNet network. The authors concluded that the deep learning method was better than the SIFT encoded method.

In our previous study~\cite{Kerkech2018}, a new method for detecting \textit{Esca} disease in UAV RGB images was proposed. The method uses LeNet5 CNN architecture and good results were obtained achieving $95\%$ disease detection accuracy in the visible range.

To the best of our knowledge, no research has been conducted so far on the combination of visible and infrared UAV images for vine disease detection by a deep learning segmentation approach.\\

\section{Study areas and materials}
\label{AreasAndMaterials}
\subsection{Study areas}
\begin{figure*}[!h]
	\centering{\includegraphics[scale=0.85]{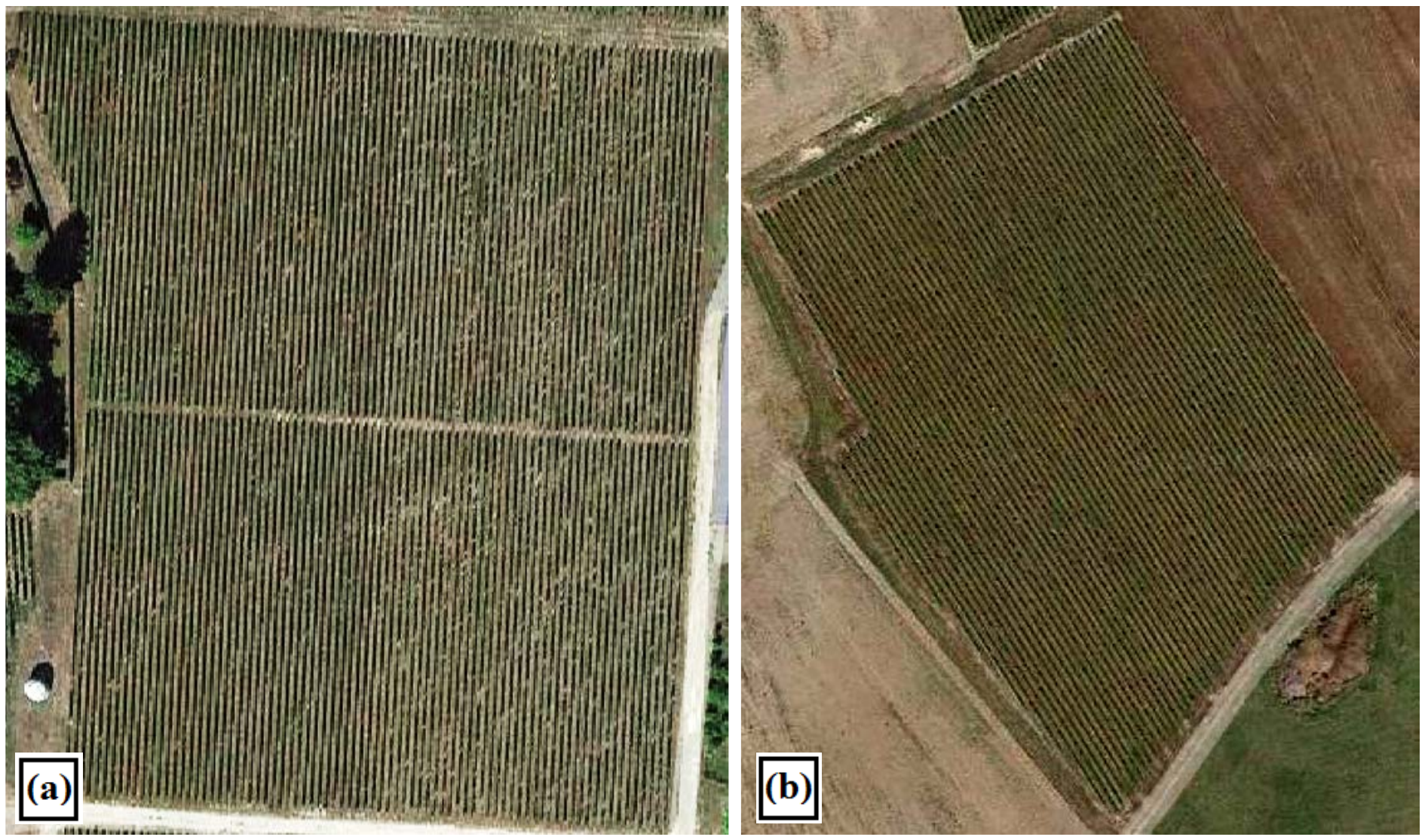}}
	\caption{\label{PlotSat}The study vineyard seen by satellite. The plot~(P1) in picture~(a) and The plot~(P2) in picture~(b).}	
\end{figure*}

\begin{figure*}[!h]
	\centering{\includegraphics[scale=0.50]{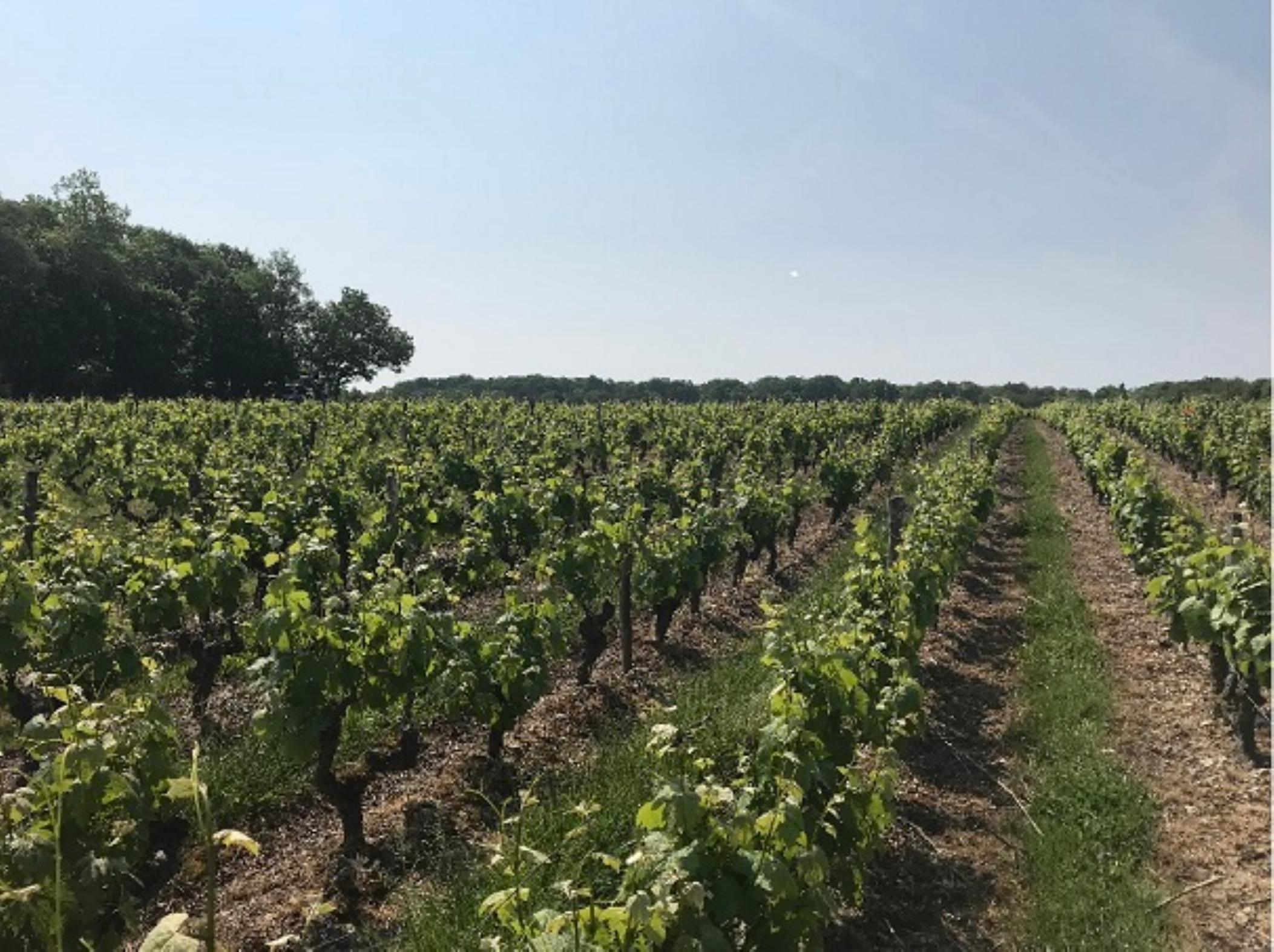}}
	\caption{\label{PlotGround}The plot (P1) seen from the ground.}	
\end{figure*}

This study is carried out on two parcels of vines located in the Center Val de Loire region in France. The first plot (P1) can be seen in Figure~\ref{PlotSat}a and the second one (P2) in Figure~\ref{PlotSat}b. P1 and P2 are at an altitude of $110$ and $114$ meters respectively, a surface of $1.8$ hectares for P1 and $1.5$ hectares for P2, and are positioned on a silty sand soil for P1 and sandy loam soil for P2. The ground of P1 is slightly inclined ($7\%$ of slope) (Figure~\ref{PlotGround}) and flat for P2. The Center Val de Loire region is characterized by a moderate temperature variation between $1$ to $26^\circ C$, and the average annual rainfall reached $700$ mm. The Table~\ref{TabPlotDescription} gives more details about the plots.

In order to carry out this study and get a healthy and diseased samples, a part of the P1 plot was treated with phytosanitary products (to protect the vine against diseases), and the other part remained untreated in order to allow the development of the disease, and getting the disease and healthy samples of vine. During this time, late blight disease was spread to all untreated areas of the plot. The P2 plot is treated globally against diseases (totally healthy), in our study, it will be used for a qualitative validation.

\begin{table*}[!h]
	\normalsize
	\renewcommand{\arraystretch}{1.25}
	\caption{\label{TabPlotDescription} The study vineyard description.}	
	\begin{center}
		\setlength{\doublerulesep}{0pt}
		\begin{tabular}{C{4cm}  C{4cm}  C{4cm} }
			\hline\hline\hline
			Types of information   &   Description (P1) & Description (P2) \\ \hline
			Surface	&   $ 1.8 $ hectares & $ 1.5 $ hectares\\ 
			Altitude 	&   $ 110 $ meter & $ 114 $ meter \\ 
			Planting year	&  $ 1976 $  &  $ 1991 $ \\ 			    	
			Variety	&  Malbec & Sauvignon \\ 
			Vine tree spacing	&  $ 1 $ meter  & $ 1 $ meter\\ 
			Interline distance 	&  $ 1.5 $  meter & $ 1.5 $  meter\\ 
			Slope 	&   $ 7\% $ & $ 0\% $ \\ 
			Exposure	&  North-South & Northwester-Southeaster\\ 
			Coarse fraction	&   $ 25\% $ & $ 17\% $ \\		    				    	
			Soil Organic Carbon 	&   $ 4.46\% $ & 3.94\% \\ 
			Soil type	&   Silty-Sand &  Sandy-Loam \\ 	\hline\hline\hline		    	
			
		\end{tabular}
	\end{center}
\end{table*}

\subsection{Materials}
\begin{figure}[!h]
	\begin{center}
		\centering{\includegraphics[scale=0.65]{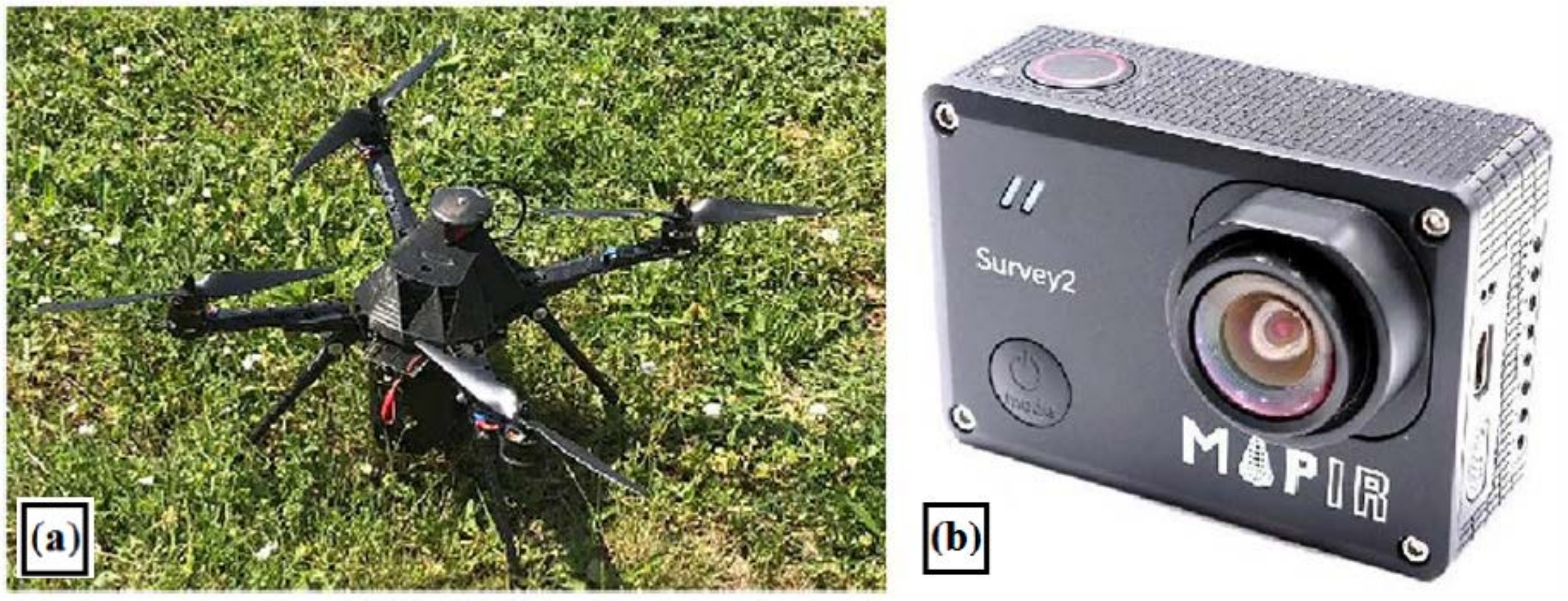}}
		\caption{\label{Materials} The acquisition materials used in this study. The quadcopter UAV drone~(a) and the high-resolution Survey2 sensor~(b).}	
	\end{center}
\end{figure}

The UAV used in the data acquisition process is a Quadcopter drone (Figure~\ref{Materials}a) which was manufactured by Scanopy. This drone embeds two cameras type sensors MAPIR Survey2 (Figure~\ref{Materials}b). The first is a visible light sensor (RGB) set to automatic lighting, and the second is an infrared light sensor (Near Infrared (NIR), Red and Normalized Difference Vegetation Index (NDVI)). For this one, the wavelength of its near infrared is 850 nm. Both image sensors have a high-resolution of $16$ megapixels (a size of $4608~\times~3456$).

The data acquisition is carried out by the drone, which flies over the plot at an altitude of $25$ meters and with an average speed of $10$ $km/h$. At this altitude, the ground resolution is $1cm^{2}/pixel$. Each $2$ seconds an image is taken automatically and without the drone stop, each image taken has a overlap over $70\%$ with the previous image. The drone has an average energy autonomy of $20$ minutes. The climatic conditions of the acquisition are moderate, which means low winds and optimal lighting (the hours of acquisition are between $11$:$30$ and $13$:$30$ to avoid the shadow of the vinerow). The acquisition was made in summer $2018$.

\section{Methods}
\label{MaterialsAndMethods}
Using the multispectral and a standard RGB images, automatic processing and analysis methods were developed to extract relevant information and correlate it with the ground truth results. Deep learning segmentation was applied on the two types of images to automatically delineate different regions (healthy, symptomatic, etc.). This generates a disease map of the vineyard comprising different segmented regions, which can be used to monitor the vineyards.

\begin{figure}[!h]
	\centering{\includegraphics[width=\textwidth]{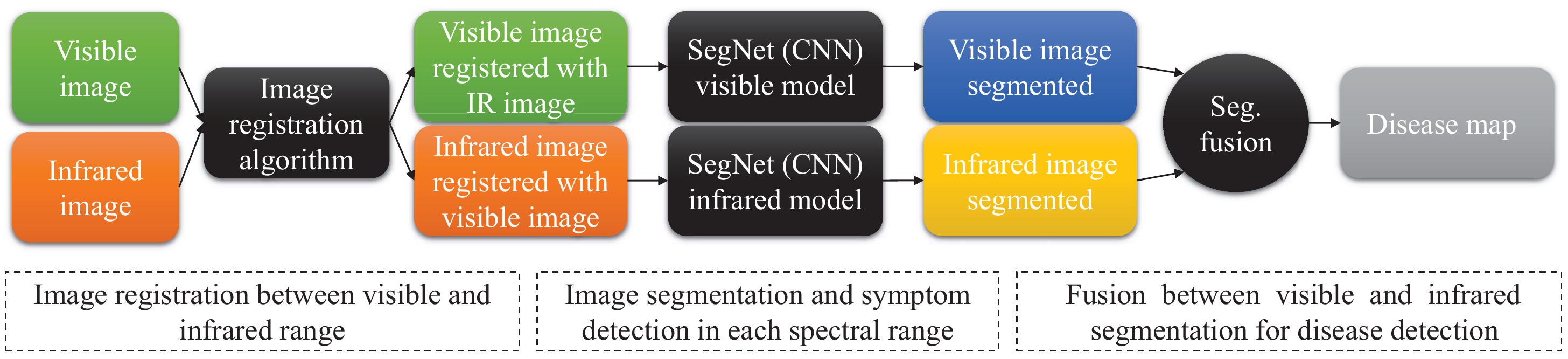}}
	\caption{\label{FigOverview} Overview of disease detection system in grapevine fields.}	
\end{figure}

The method comprises three main steps (see Figure~\ref {FigOverview}). The first one consists in image registration between the images acquired in the visible and infrared range. This step is essential for the third step, as it enables the pixel-wise superposition of the two images and thus allows segmentation fusion. Once the registration of the two images has been performed, the next stage consists in segmenting the plot in the visible and infrared range using the SegNet architecture. The two segmented images are merged in the third step, to generate a disease map.

\subsection{Image registration}
The objective of the registration algorithm is to realign and geometrically correct the shift~\cite{Szeliski2006} between the visible and infrared images at the pixel-wise level. Generally, the UAV image is partially distorted due to the UAV vibrations, read in rolling shutter mode of the sensors and optics. Therefore, the rigid alignment model (translation and rotation) is not appropriate in our case. Hence, the proposed algorithm uses the non-rigid model. However, even if the registration model is correct it is not sufficient to align the two types of images perfectly. Moreover, matching image points between the visible and infrared bands is difficult, since the key points do not necessarily have the same visual properties in the two spectral bands. In order to improve the accuracy of the image alignment, an iterative process based on minimization of the registration error was implemented.

The registration algorithm proposed in this study is based on the Accelerated-KAZE (AKAZE) algorithm~\cite{Alcantarilla}. AKAZE is an algorithm used in computer vision for detecting objects or similarities in two images. Its principle is comparable to that of the Scale Invariant Feature Transform (SIFT)~\cite{Lowe2004}, Speeded Up Robust Features (SURF)~\cite{Bay}, Features from Accelerated Segment Test (FAST), Binary Robust Independent Elementary Features (BRIEF), Oriented FAST and Rotated BRIEF (ORB)~\cite{Rublee2011}, KAZE~\cite{Fern2012} algorithms. However, AKAZE is much more efficient in detection robustness, description and in the speed of calculation as it was created with high-performance algorithms in a pyramidal framework comparable to other algorithms of the same type. Even if AKAZE follows the same pyramidal steps as the other algorithms, the method used is very different. AKAZE integrates Fast Explicit Diffusion (FED)~\cite{Grewenig2010} systems for accelerated feature detection in a non-linear scale space. Moreover, a modified version of Local Difference Binary (M-LDB)~\cite{Alcantarilla} has been integrated into AKAZE. Unlike the old version, this modified version of LDB~\cite{Yang2014} is a rotationally invariant, scaled descriptor and can exploit gradient information from the non-linear scale space.

\begin{figure}[!h]
	\centering{\includegraphics[width=14cm]{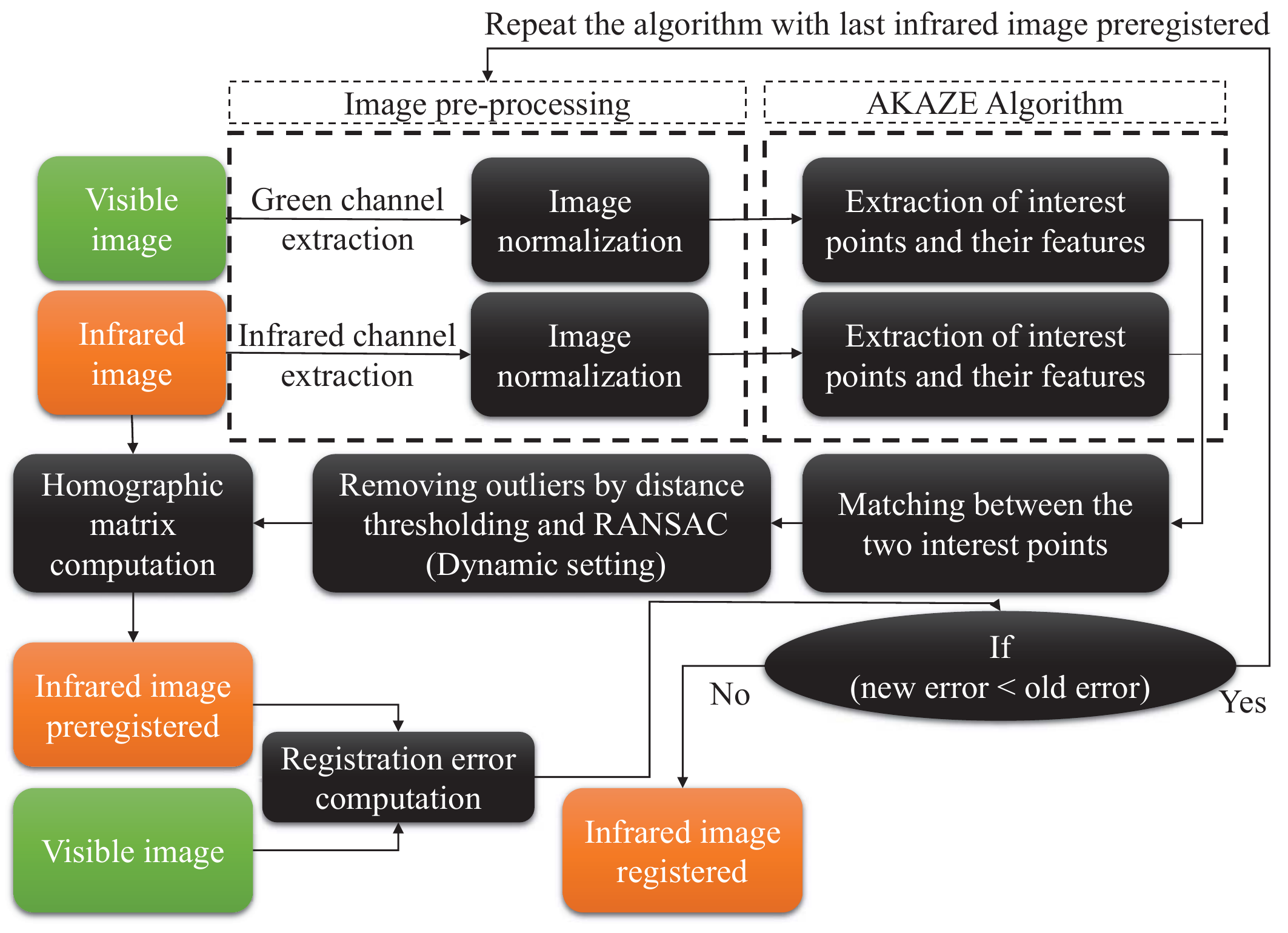}}
	\caption{\label{ImRegistrationSch} Proposed method for non-rigid registration of multimodal (visible and infrared) image.}	
\end{figure}

The proposed registration method is shown in Figure~\ref{ImRegistrationSch}. The first step is to extract the green channel (G) from the visible image, and the near infrared channel (NIR) from the infrared image (these channels were selected for their vegetation texture information). By using the normalization equation~(\ref{I_Normalized}), the second phase normalizes the two spectral channels to improve their contrast. The third step is to extract the points of interest and calculate their features from the two channels (by the AKAZE algorithm). Based on the features of the interest points, the fourth stage is to map each point of interest extracted in the G channel to the corresponding point in the NIR channel. In the fifth step, to eliminate some outliers, a first thresholding algorithm performs a preselection of inlier matching points, then a final selection is performed by the RANdom SAmple Consensus (RANSAC) algorithm~\cite{Fischler1981}. To obtain the best setting for algorithms that contribute to removing the outliers, a dynamic setting (by threshold variation) method based on homographic matrix analysis is integrated into the registration system. This dynamic method (dynamic threshold) very significantly reduces the number of registration failures. Note that X is the coordinates of the source infrared image (x, y), X' is the coordinates of the registered infrared image (x', y') and H is the homographic matrix~\ref{MatHomographic}. In order to avoid the problems that lead to misregistration, a dynamic threshold regulation of the RANSAC algorithm was used, and the distance-based algorithm. For each given value, a homographic matrix is estimated. The viability of this matrix is then tested by equation~\ref{EqReg}. This test is performed by projecting the end coordinates of the source image into the new space of the registered image. If the result found is coherent (the distance between the old coordinates and the new coordinates does not exceed a certain threshold), this implies the end of the dynamic adjustment procedure, and the matrix tested is used to register the image. Otherwise, a new setting is made to repeat the same procedure. Once this step is finished, the pre-registered infrared image can be obtained by the homographic matrix validated by the dynamic adjustment procedure.

In order to reduce the registration error, an iterative method was implemented. After the pre-registration has been completed, an iterative phase starts from the result obtained, calculates the error by the Root Mean Squared Error (RMSE)~\cite{Ross2019} between matched points which are calculated on the X coordinates with equation (\ref{RMSE_x}), and on the Y coordinates with equation (\ref{RMSE_y}), and then calculates their module by equation (\ref{RMSE}). If the error decreases, another iteration is performed, if not, the iterative process stops and the best result, which is the result with the minimum RMSE, is kept.

\begin{equation}
\label{I_Normalized}
\begin{array}{c}
$$I_{Normalized} = 255 \times \frac{I-min(I)}{max(I)-min(I)}$$
\end{array}
\end{equation}

where \textit{$I_{Normalized}$} is the normalized image of the source image $I$ (visible or infrared), $min(I)$ and $max(I)$ are respectively the minimum and maximum grey level value of the source image $I$.

\begin{equation}
\label{MatHomographic}	
H = 
\begin{pmatrix} 
$$1+h_{00}$$ & $$h_{01}$$ & $$h_{02}$$\\ 
$$h_{10}$$ & $$1+h_{11}$$ & $$h_{12}$$\\ 
$$h_{20}$$ & $$h_{21}$$ & $$1$$\\ 		
\end{pmatrix}
\end{equation}

\begin{equation}
\label{EqReg}	
\begin{array}{c}
$$X' = H X$$
\end{array}
\end{equation}

%

where $(1+h_{00})$ and $(1+h_{11})$ are stable scale factors respectively in the X and Y direction only. $h_{01}$ and $h_{10}$ are scale factors respectively in the X direction relative to the Y distance from the origin and Y direction relative to the X distance from the origin. $h_{02}$ and $h_{12}$ are respectively translation in the X and Y direction.$h_{20}$ and $h_{21}$ are relative scale factors X and Y respectively as a function of X and Y.

\begin{equation}
\label{RMSE_x}
\begin{array}{c}
$$RMSE_{x} = \sqrt{\frac{\sum_{i=0}^{N}(x_{VIS i}-x_{IR i})^2}{N}}$$
\end{array}
\end{equation}

\begin{equation}
\label{RMSE_y}
\begin{array}{c}
$$RMSE_{y} = \sqrt{\frac{\sum_{i=0}^{N}(y_{VIS i}-y_{IR i})^2}{N}}$$
\end{array}
\end{equation}

\begin{equation}
\label{RMSE}
\begin{array}{c}
$$RMSE = \sqrt{RMSE^{2}_{x}+RMSE^{2}_{y}}$$
\end{array}
\end{equation}	

where $(x_{VIS i}$, $y_{VIS i})$ and $(x_{IR i}$, $y_{IR i})$ are respectively the "i" th coordinates of correspondence between the visible and infrared images. N is the number of matches found between the visible and infrared images.

\subsection{Segmentation and fusion}
This subsection presents the visible and infrared image segmentation system, the deep learning architecture used in this process, the labelling and learning method, and finally, the overall operation.

\subsubsection{Deep learning segmentation}
\begin{figure}[!h]
	\centering{\includegraphics[width=\textwidth]{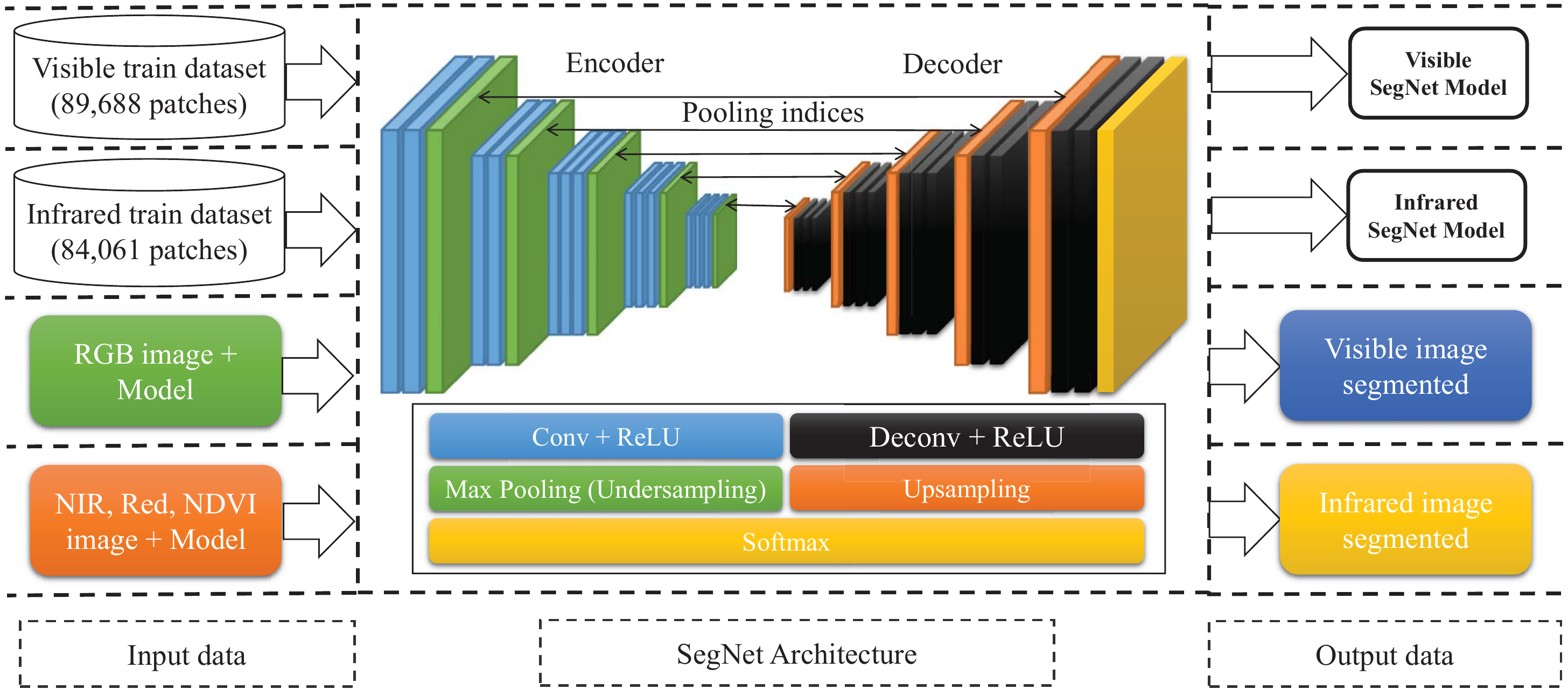}}
	\caption{\label{SegNetArch} Visible and infrared image modeling and segmentation system.}	
\end{figure}

The CNN architecture has been very successful in the pattern recognition and computer vision fields. Since then, there has been a continuous evolution of CNN architectures. The new architectures have become deeper, but also new types of architectures have emerged that directly segment an image, such as the SegNet~\cite{Badrinarayanan} architecture, used in the present study. To segment an image, the SegNet architecture (Figure~\ref{SegNetArch}) operates through two opposite phases, an information encoding phase and a decoding and classification phase (Table~\ref{TabSegNetSetting}). The encoding phase is in fact a classic CNN architecture, usually with a VGG-16~\cite{Simonyan2014} architecture. The coding phase consists of three types of processing, namely convolution layers, ReLU layers (ReLU is an activation function for removing the negative values that result from convolution and deconvolution. It is commonly used in deep learning networks, because it performs better than other activation functions) and MaxPooling layers (non linear sub-sampling function). Decoding consists of the same types of processing except for the convolution layers which are replaced by deconvolution layers, and the MaxPooling layers which are replaced by Upsampling layers. It is in the decoding part of the network that the segmented image is formed, until the final decoding layer is reached. At this level, a pixel-wise segmentation is performed by the Softmax function.

In the present study, generation of the disease map of a vineyard field is considered as a four-class segmentation problem in the visible and infrared range. The objective is to build a SegNet model capable of differentiating between shadow, ground, healthy and symptomatic (visible and infrared) classes in the two spectral bands. The distinction between each class is mainly based on variations in color, texture, spectral information and spatial relative position of each class. This important information must be extracted by the SegNet network during the encoding phase, also called the feature extraction phase, then rebuilt and segmented by the decoding phase.

\begin{table}[!h]
	\normalsize
	\renewcommand{\arraystretch}{1.25}
	\caption{\label{TabSegNetSetting} SegNet setting.}	
	\begin{center}	
		\begin{tabular}{|c | c | c | c | c|} 
			\hline
			Phase & Number & Layer type & Filter size & Number of features maps \\
			\hline
			\multirow{13}{*}{Encoder} 
			&\multirow{2}{*}{1} & Conv1-1 & $3 \times 3$ & $64$\\
			\cline{3-5}
			&& Conv1-2 & $3 \times 3$ & $64$\\
			\cline{2-5}
			&\multirow{2}{*}{2} & Conv2-1 & $3 \times 3$ & $128$\\
			\cline{3-5}
			&& Conv2-2 & $3 \times 3$ & $128$\\
			\cline{2-5}
			&\multirow{3}{*}{3} & Conv3-1 & $3 \times 3$ & $256$\\
			\cline{3-5}
			&& Conv3-2 & $3 \times 3$ & $256$\\
			\cline{3-5}
			&& Conv3-3 & $3 \times 3$ & $256$\\
			\cline{2-5}
			
			&\multirow{3}{*}{4} & Conv4-1 & $3 \times 3$ & $512$\\
			\cline{3-5}
			&& Conv4-2 & $3 \times 3$ & $512$\\
			\cline{3-5}
			&& Conv4-3 & $3 \times 3$ & $512$\\
			\cline{2-5}
			
			&\multirow{3}{*}{5} & Conv5-1 & $3 \times 3$ & $512$\\
			\cline{3-5}
			&& Conv5-2 & $3 \times 3$ & $512$\\
			\cline{3-5}
			&& Conv5-3 & $3 \times 3$ & $512$\\
			\hline
			\multirow{13}{*}{Decoder} 
			&\multirow{3}{*}{5} & Deconv5-3 & $3 \times 3$ & $512$\\
			\cline{3-5}
			&& Deconv5-2 & $3 \times 3$ & $512$\\
			\cline{3-5}
			&& Deconv5-1 & $3 \times 3$ & $512$\\
			\cline{2-5}	
			&\multirow{3}{*}{4} & Deconv4-3 & $3 \times 3$ & $512$\\
			\cline{3-5}
			&& Deconv4-2 & $3 \times 3$ & $512$\\
			\cline{3-5}
			&& Deconv4-1 & $3 \times 3$ & $512$\\
			\cline{2-5}
			&\multirow{3}{*}{3} & Deconv3-3 & $3 \times 3$ & $256$\\
			\cline{3-5}
			&& Deconv3-2 & $3 \times 3$ & $256$\\
			\cline{3-5}
			&& Deconv3-1 & $3 \times 3$ & $256$\\
			\cline{2-5}
			&\multirow{2}{*}{2} & Deconv2-2 & $3 \times 3$ & $128$\\
			\cline{3-5}
			&& Deconv2-1 & $3 \times 3$ & $128$\\
			\cline{2-5}
			&\multirow{2}{*}{1} & Deconv1-2 & $3 \times 3$ & $64$\\
			\cline{3-5}
			&& Deconv1-1 & $3 \times 3$ & $64$\\
			\hline
		\end{tabular}
	\end{center}
\end{table}

\subsubsection{Fusion of multimodal image segmentation}
The fusion of segmentations is performed in order to obtain a disease map with more robust results. To generate a disease map, each pixel of the image segmented in the visible range is compared with the pixel of the same position in the infrared range. Here, three main cases are considered. The first one is that the two pixels represent the symptomatic class. In this case, the result is symptom intersection class. The second case is when the pixel is classified as symptomatic in the infrared range, and healthy in the visible range. In this case, the resulting class is symptomatic infrared (This may be a case where the disease has not yet affected the visible range by discoloration of the leaves). The third case is when the pixel is classified as healthy in the infrared range, and symptomatic in the visible range. The resulting class is visible symptomatic. 

To evaluate the disease map, two cases are evaluated. The first one, is the case described in the previous paragraph, it is named fusion by intersection and symbolized "Fusion~AND", the AND operator means the symptom is considered to be detected if it is present in both visible and infrared images. The second case is named fusion by union and is symbolized "Fusion~OR". As are named, this case unites visible and infrared ranges detections with or operator.

\subsection{Dataset}
The dataset was composed of visible and infrared range images. In our case, the visible sensor was used to detect the presence or absence of chlorophyll in the crop, i.e. to detect any anomalies in the vegetation in relation to its discoloration. The acquisition wavelength of the infrared sensor used here is $850$ nm. Unlike visible wavelengths, this wavelength was chosen for its sensitivity to changes in the different states of vegetation. Indeed, due to the high reflectance generated by vegetation, near infrared waves are relevant for plant analysis.

\begin{figure}
	\centering{\includegraphics[width=\textwidth]{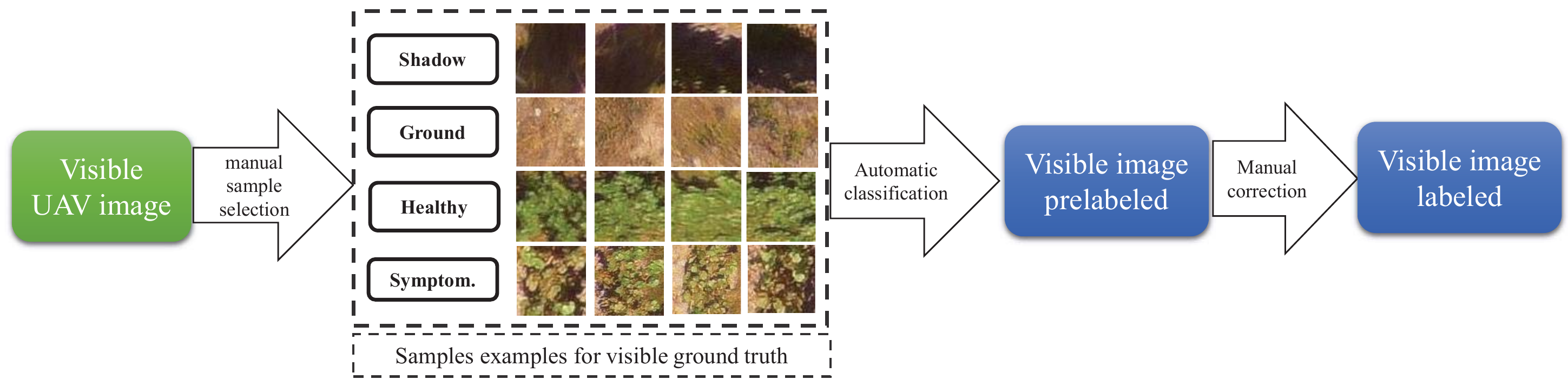}}
	\caption{\label{VIS_GT} Semi-automatic ground truth for the visible image.}	
\end{figure}

\begin{figure}
	\centering{\includegraphics[width=\textwidth]{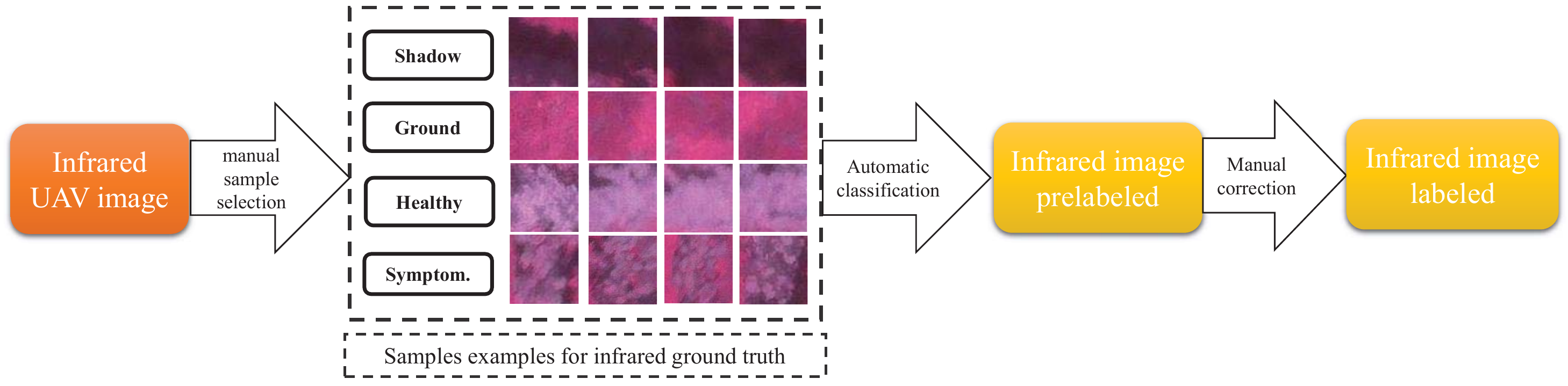}}
	\caption{\label{NIR_GT} Semi-automatic ground truth for the infrared image.}	
\end{figure}

The visible (Figure~\ref{VIS_GT}) and infrared (Figure~\ref{NIR_GT}) datasets were labelled based on four dominant classes in the visible and infrared range, fusion by intersection, and fusion by union. The first class represents the shaded areas in the vine and on the ground (all the dark areas). This class does not reflect light, which means that no relevant information can be drawn from it. The second class represents the ground; it can be an area of weeds or an area of any type of soil. The third class represents the healthy vegetation of the vine. This class has a green color in the visible range and has a high reflectance level compared to the previous classes in the infrared range. Finally, the fourth class represents the symptomatic vegetation of the vine. This class is generally yellow or brown color in the visible range (in the case of an advanced symptom). This color results from a problem with chlorophyll production. In the infrared range, the symptomatic class is a variation in reflectance between a leaf and its neighbours. Visually this gives a special texture characterized by a significant variation in reflectance, with which this class can be distinguished from the others. After merging the segmented visible and infrared images, a class called "symptomatic intersection" is created when symptoms are detected in the same visible and infrared area, and "symptomatic union" is created for unified the two symptoms class.

By using data augmentation methods (described in the following sub-section), a dataset of $105,515$ and $98,895$ patches ($360~\times~480$ pixels for each patch) was generated for the visible and infrared range respectively. This dataset was used to train and validate the SegNet model. Among the dataset patches, $85\%$ was randomly selected for training and the remain ($15\%$) was used for validation. To evaluate and test the SegNet model, another area (the second area) at two different dates was used to test the visible and infrared model, then the fusion algorithm.

\subsection{Data augmentation}
Due to the huge amount of data required to train a deep learning network, the lack of data, and the difficulty of labelling images, several methods of data augmentation~\cite{Dellana2016} were used. First, some images acquired by UAV with a size of $4608~\times~3456$ pixels were labelled by a semi-automatic method. Then each of these images underwent automatic data augmentation to generate $360~\times~480$ pixels labelled patches, which were used in SegNet network learning. 

In this process, several data augmentation techniques were combined. The first transformation consisted in horizontal shift by an overlap of $50\%$ to enable the network to learn the transition areas. Rotation was also used, since it enables the network to learn the different orientations of the vinerows. Several rotation values ranging between $0^\circ$ and $180^\circ$ with a step of $30^\circ$ were used. The third transformation consisted in under and over sampling (scale change) in order to enable the network to learn that the thickness of the vine rows could change, but that it would not affect the classification in the case of a scale change, such as a change in altitude of the UAV for image acquisition. For that purpose, a sub-sampling between $0.5$~to~$1$ of the real scale, and an oversampling between $1$~to~$1.5$ was performed with a step of $0.25$. The fourth technique of data augmentation involved modification of the brightness to make the network insensitive to light levels. As the image acquisition was done outdoors, the brightness parameters are uncontrollable because of changes in the weather. Thus, coefficients between $0.8$ to $1.2$ with a step of $0.1$ were multiplied with the grey levels of the images to generate dark, normal, and bright effects.\\

\section{Experimentation}
\label{Experimentation}
This section details the experiments, testing, validation and interpretation of the results. The algorithms were developed under the Python $ 2.7 $ development environment by using the Tensorflow $ 1.8.0 $, NumPy $ 1.16.2 $ and OpenCV $ 3.0.0 $ libraries. To run and evaluate the runtime of our algorithms, we used the following hardware; an Intel Xeon (R) W-2123 (a) $ 3.60 $ GHz$\times$$ 8 $ processor (CPU) with $ 32 $ GB of RAM, and a graphics processing unit (GPU) NVidia GTX 1080Ti with an internal RAM of $11$ GB under the Linux operating system Ubuntu $ 16.04~LTS $ ($ 64 $ bits).

The experiments section is divided into two subsections. The first one concerns the visible and infrared image registration, and the second one details the segmentation and fusion steps.

\subsection{Evaluations of image registration algorithm}
Performance measurement was carried out by computing the RMSE Eq.~\ref{RMSE_x},~\ref{RMSE_y} and~\ref{RMSE} between the points of interest matched, in pixel units. RMSE provides information on the geometric correction and the shift between the visible and infrared images. 

\begin{table}[h!]
	\normalsize
	\renewcommand{\arraystretch}{1.25}
	\caption{\label{TabRegistration}Statistical performance results for standard and optimized registration for a dataset of 150 images. Mean, Min and Max are the average, minimum and maximum number of the statistic results, respectively.}	
	\begin{center}	
		\setlength{\doublerulesep}{0pt}
		\begin{tabular}{|c | c c c | c c c|}
			\hline
			\multirow{2}{*}{\backslashbox{Measure}{Methods}} & \multicolumn{3}{c}{Standard registration} & \multicolumn{3}{|c|}{Optimized registration}\\
			\cline{2-7}
			& Mean & Min & Max & Mean & Min & Max \\
			\hline
			RMSE "Pixels" & $3.29 \pm 1.57$ & $1.13$ & $9.75$ & $\textbf{2.43} \pm \textbf{1.26}$ & $\textbf{0.87}$ & $\textbf{9.02}$\\
			\hline
			Runtime "Seconds" & $\textbf{92} \pm \textbf{19}$ & $\textbf{49}$ & $\textbf{129}$ & $139 \pm 40$&	$67$ & $238$\\
			\hline
			Number of iterations &	- &	- &	- & $3.12 \pm 1.48$ & $1$ & $7$\\
			\hline
		\end{tabular}
	\end{center}
\end{table}

\begin{figure}[h!]
	\centering{\includegraphics[width=\textwidth]{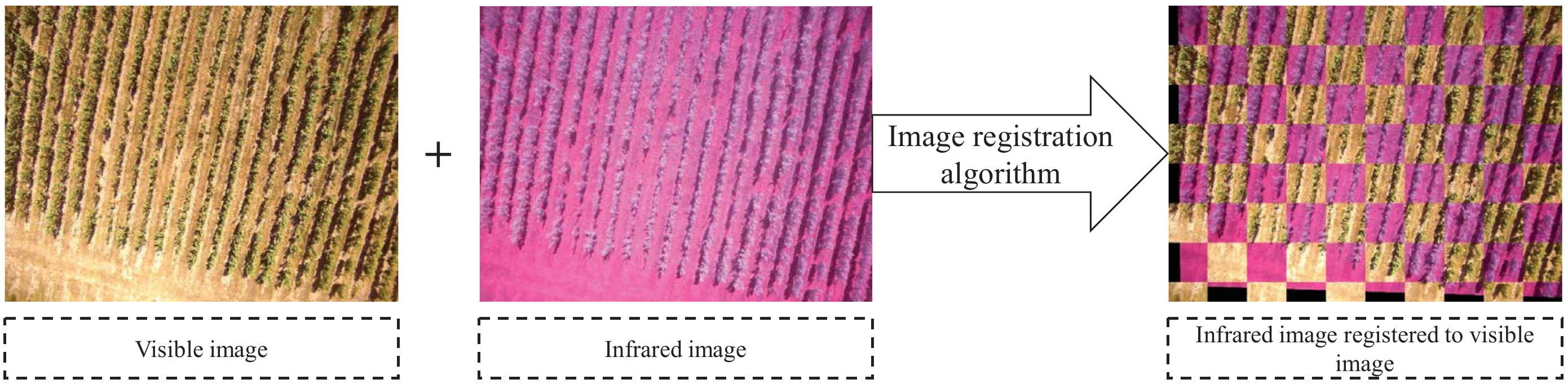}}
	\caption{\label{ImRegistrationExemple} Result on image registration.}	
\end{figure}

\begin{figure*}[h!]
	\centering{\includegraphics[width=15cm]{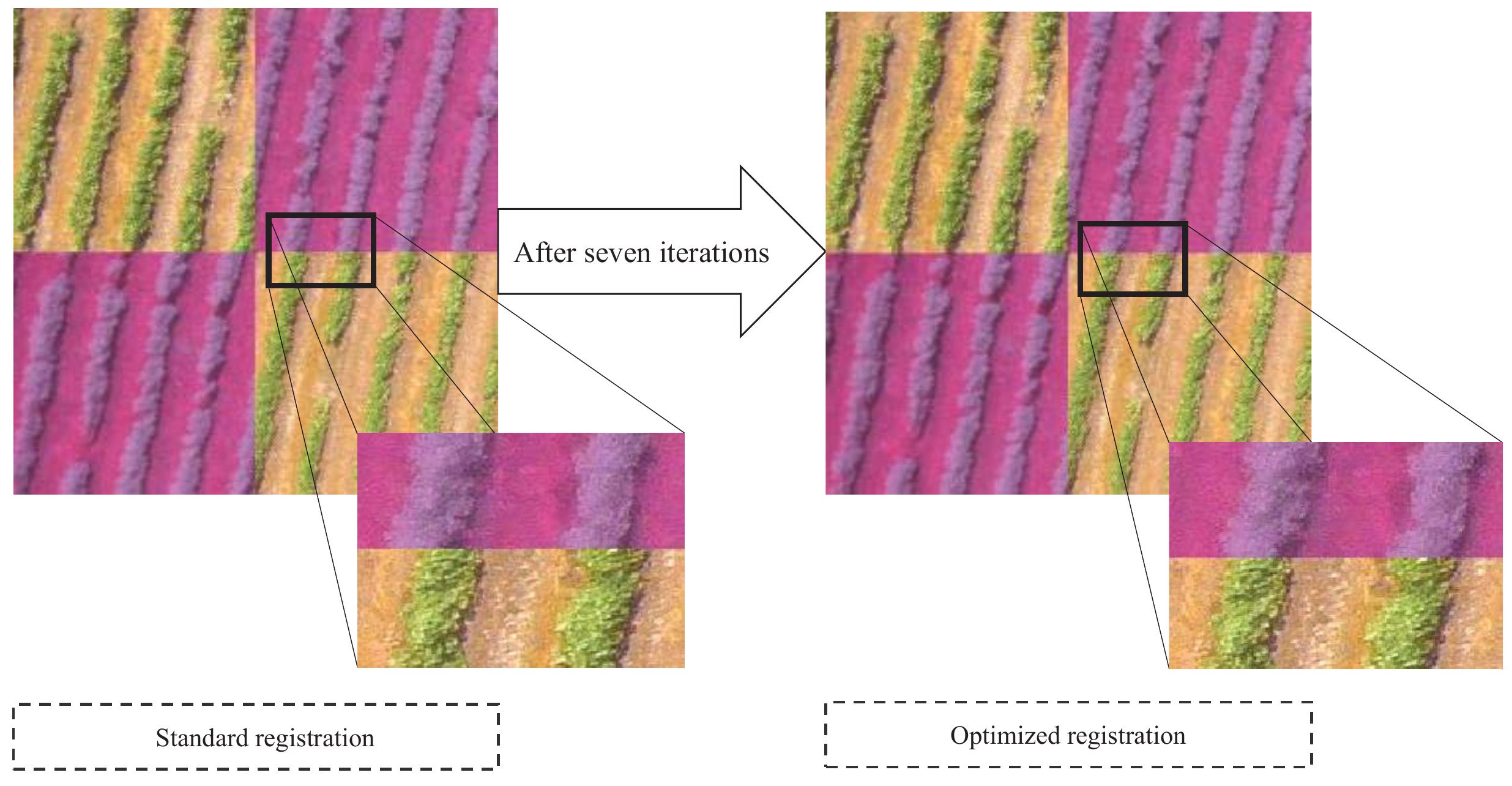}}
	\caption{\label{ImRegistrationOptimExemple} Correction of the shift by seven iterations.}	
\end{figure*}

\subsection{Experiment on image segmentation and fusion}
\subsubsection{Data labelling}
Due to the large amount of data and the difficulty of achieving accurate labelling, which must be provided for learning, the data labelling procedure was performed by a semi-automatic method. First, an automatic step was performed by a CNN LeNet5 network for pre-labelling. Then, the manual labelling was reinforced by the ground truth provided by technicians in the field.

Two patch datasets were created for visible and infrared images, the patch size is $32 \times 32$, organized into $4$ classes (shadow, ground, healthy and symptomatic class). Each dataset contains $70,560$ patches ($17,640$ samples for each class, among these samples, there are $14,994$ samples for training and $2,646$ samples for validation). Then, two CNN LeNet5 models were created from these two datasets (visible and infrared). The sliding window method was used to perform automatic labelling on UAV images with a size of $4608~\times~3456$ pixels. To obtain the best possible accuracy, the sliding window was set with a displacement step of $2~\times~2$. The automatic labelling operation required just over an hour and a half of runtime to label only a single UAV image. The manual correction, performed by the free software "Paint.Net", consists in correcting possible classification errors generated by automatic labelling and adding the symptomatic class. Indeed, technicians observed vines at the ground and reported all the diseased ones. By referring to the ground truth, each vine was referenced with its coordinates in the field. Then, this information was used to label the areal images.

\subsubsection{Learning and testing procedure}	

First, the labelled visible and infrared datasets were used in the learning phase. During this phase, the SegNet network uses the patches sized $360~\times~480$ pixels, with their labels. This learning phase was performed for visible and infrared ranges separately. The number of iterations was set to $100,000$, each batch was composed of $ 5 $ patches selected randomly.

The SegNet models generated were used for testing and evaluation. The segmented images were compared with the ground truth to estimate the segmentation accuracy. Due to the large size of the UAV image ($4608~\times~3456$ pixels), images were divided into non overlapping blocks, with a size of $360~\times~480$ pixels. Segmented blocks were stitched together to retrieve the original size.

\subsubsection{Performance measurement}
The segmentation performance was measured by two methods. The first one (presented in Table.\ref{TabSegResult}) is based on the leaf-level (pixel-wise) computation of the Recall~(eq.\ref{Recall}), precision~(eq.\ref{Precision}), F1-Score~(eq.\ref{F1Score})/Dice coefficient~(eq.\ref{Dice}) and Accuracy~(eq.\ref{ACC}) for each class (shadow, ground, healthy and symptomatic). The second evaluation is based on the grapevine-level (presented in Table.\ref{TabDetectionResult}). Indeed, the segmentation measurement at pixel-wise do not provide an information about the detection at the grapevine-level, it indicates a given grapevine is infected or not. This measurement uses a sliding window with a size of $64\times64$ pixels (corresponding to average size of a grapevine in the studied plots). Inside this window, only the dominant class in the ground truth is evaluated. If there is a match between the ground truth and the SegNet estimation, it is considered as true positive, otherwise it is false positive. At the end of this process, the same measurement is computed. 

\begin{equation}
\label{Recall}
\begin{array}{c}
$$Recall = \frac{TP}{TP+FN}$$
\end{array}
\end{equation}

\begin{equation}
\label{Precision}
\begin{array}{c}
$$Precision = \frac{TP}{TP+FP}$$
\end{array}
\end{equation}

\begin{equation}
\label{F1Score}
\begin{array}{c}
$$F1-Score = 2 \frac{Recall \times Precision}{Recall+Precision} = \frac{2TP}{FP+2TP+FN}$$
\end{array}
\end{equation}

\begin{equation}
\label{Dice}
\begin{array}{c}
$$Dice = \frac{2 |X \cap Y|}{|X|+|Y|} = \frac{2(TP)}{(FP+TP)+(TP+FN)} = \frac{2TP}{FP+2TP+FN}$$
\end{array}
\end{equation}

\begin{equation}
\label{ACC}
\begin{array}{c}
$$Accuracy = \frac{TP+TN}{TP+TN+FP+FN}$$
\end{array}
\end{equation}

where TP, TN, FP and FN are the number of samples for "True Positive", "True Negative", "False Positive" and "False Negative". For the Dice equation, X is the set of ground truth pixels and Y is the set of pixels estimated by the SegNet classifier.
\begin{table}[!h]
	\normalsize
	\renewcommand{\arraystretch}{1.25}
	\caption{\label{TabSegResult} The leaf-level (pixel-wise) average result on two temporal tests by measuring Recall (Rec.), Precision (Pre.), F1-Score/Dice (F1/D.) and Accuracy (Acc.) on the performance of visible, infrared image segmentation and fusion (values presented in percent). Note that: "Fusion~AND" and "Fusion~OR" are the cases where their symptomatic classes are respectively the intersection and the union of the symptomatic visible and infrared classes.}	
	\setlength{\doublerulesep}{0pt}
	\begin{center}	
		\footnotesize
		\begin{tabular}{|c|c|c|c|c|c|c|c|c|c|c|c|c|c|} 
			\hline
			Class name	& \multicolumn{3}{c|}{Shadow} & \multicolumn{3}{c|}{Ground} & \multicolumn{3}{c|}{Healthy} & \multicolumn{3}{c|}{Symptomatic}& Total\\
			\hline
			Measure & Rec. & Pre. & F1/D. & Rec. & Pre. & F1/D. & Rec. & Pre. & F1/D. & Rec. & Pre. & F1/D. & Acc.\\
			\hline
			Visible & 76.31 & 87.25 & 81.05 & 91.37 & 95.95 & 93.51 & 86.86 & 66.89 & 75.31 & 80.22 & 77.99 & 78.72 & 85.13\\
			\hline
			Infrared & 84.25 & 72.25 & 77.69 & 87.74 & 91.33 & 89.42 & 73.81 & 50.18 & 58.58 & 59.02 & 85.06 & 69.55 & 78.72\\
			\hline
			Fusion~AND & 87.84 & 86.78 & 87.30 & 95.73 & 95.95 & 95.84 & 83.73 & 69.09 & 75.60 & 53.70 & 94.02 & 67.93 & 82.20\\
			\hline
			Fusion~OR & 87.84 & 86.78 & 87.30 & 95.73 & 95.95 & 95.84 & 82.12 & 72.30 & 76.55 & 84.07 & 90.47 & 87.12 & 90.23\\
			\hline			
		\end{tabular}
	\end{center}
\end{table}

\begin{table}[!h]
	\normalsize
	\renewcommand{\arraystretch}{1.25}
	\caption{\label{TabDetectionResult} The grapevine-level average result on two temporal tests by measuring Recall (Rec.), Precision (Pre.), F1-Score/Dice (F1/D.) and Accuracy (Acc.) on the performance of visible, infrared image segmentation and fusion (values presented in percent). Note that: "Fusion~AND" and "Fusion~OR" are the cases where their symptomatic classes are respectively the intersection and the union of the symptomatic visible and infrared classes.}	
	\setlength{\doublerulesep}{0pt}
	\begin{center}	
		\footnotesize
		\begin{tabular}{|c|c|c|c|c|c|c|c|c|c|c|c|c|c|} 
			\hline
			Class name	& \multicolumn{3}{c|}{Shadow} & \multicolumn{3}{c|}{Ground} & \multicolumn{3}{c|}{Healthy} & \multicolumn{3}{c|}{Symptomatic}& Total\\
			\hline
			Measure & Rec. & Pre. & F1/D. & Rec. & Pre. & F1/D. & Rec. & Pre. & F1/D. & Rec. & Pre. & F1/D. & Acc.\\
			\hline
			Visible & 94.00  & 93.42  & 93.63  & 97.39  & 97.94  & 97.66  & 95.16  & 85.20  & 89.91  & 90.15  & 92.97  & 91.50  &  94.41\\
			\hline
			Infrared & 97.53  & 79.97  & 87.55  & 97.30  & 95.41  & 96.32  & 93.72  & 69.19  & 79.19  & 70.49  & 96.92  & 81.66  &  89.16\\
			\hline
			Fusion~AND & 94.01  & 86.62  & 89.96  & 97.41  & 97.89  & 97.65  & 93.81  & 87.55  & 90.56 & 66.92  & 73.12  & 68.03  &  88.14\\
			\hline
			Fusion~OR & 94.00  & 94.03  & 93.95  & 97.39  & 97.94  & 97.66  & 93.81  & 89.65  & 91.68  & 92.91  & 92.78  & 92.81  &  95.02\\
			\hline			
		\end{tabular}
	\end{center}
\end{table}

\begin{table}
	\normalsize
	\renewcommand{\arraystretch}{1.25}
	\caption{\label{TabRuntime}Results on runtime performance (expressed in seconds) for the entire system as a function of the average runtime of standard and optimized registration.}	
	\begin{center}	
		\setlength{\doublerulesep}{0pt}
		\begin{tabular}{C{4cm} C{2cm} C{2cm} C{2cm} C{2cm}} 
			\hline\hline\hline
			Step & Registration & SegNet seg. & Fusion & Total\\
			\hline
			Standard registration & $92$ & $140 \times 2$ & $2$ & $374$\\
			Optimized registration & $139$ & $140 \times 2$ & $2$ & $421$\\
			\hline\hline\hline
		\end{tabular}
	\end{center}
\end{table}

\begin{figure}
	\centering{\includegraphics[scale=0.70]{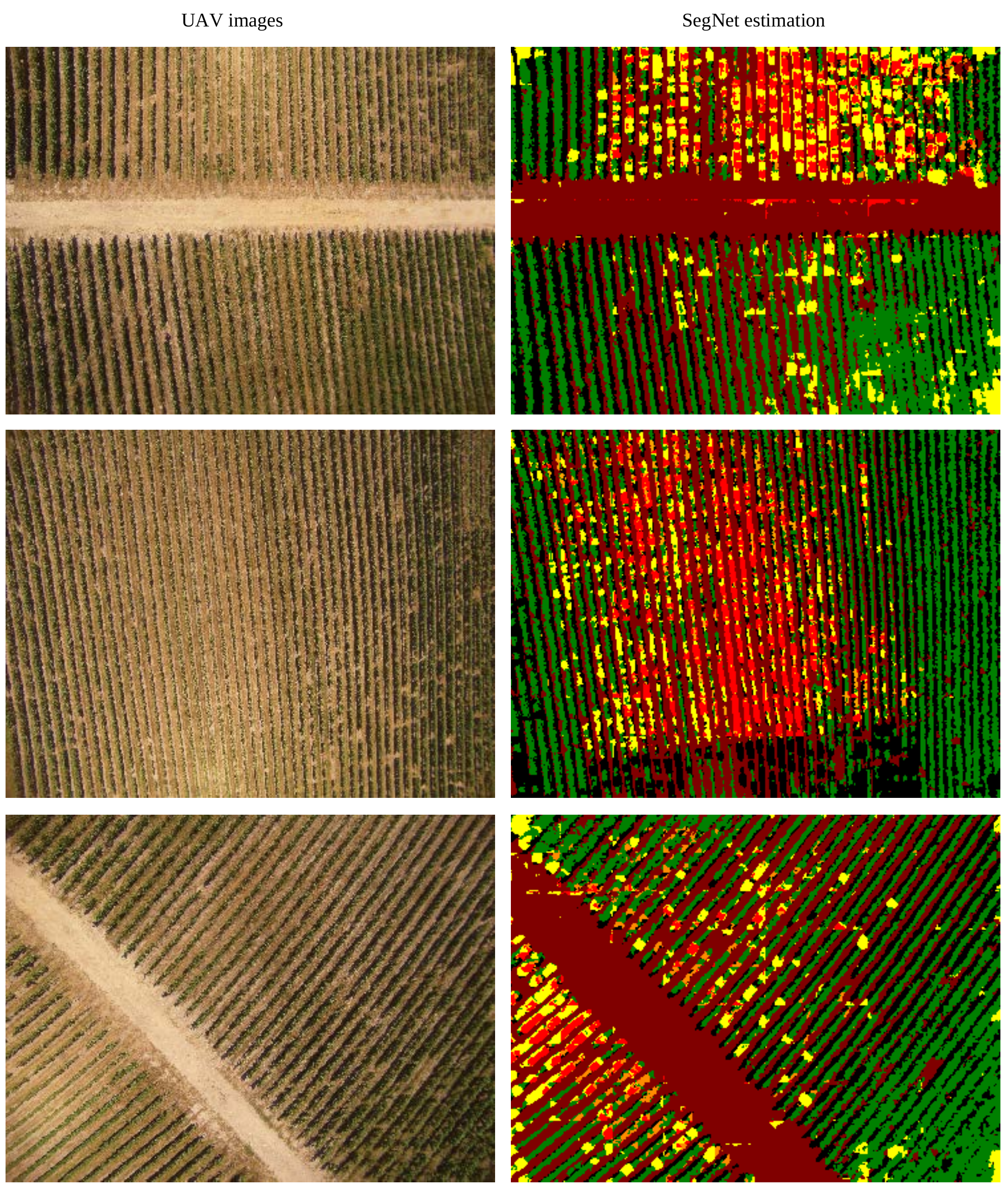}}
	\caption{\label{Qualitatifs_P1} Segmentation qualitative results by the SegNet method for the P1 vineyard. On the left, images from UAV and their right, the segmentation result of these images. The color code of the segmentation is; Black: Shadow, Brown: Ground, Green: Healthy, Yellow: Visible symptom, Orange: Infrared symptom, Red: Symptom intersection.}	
\end{figure}	

\begin{figure}
	\centering{\includegraphics[scale=0.70]{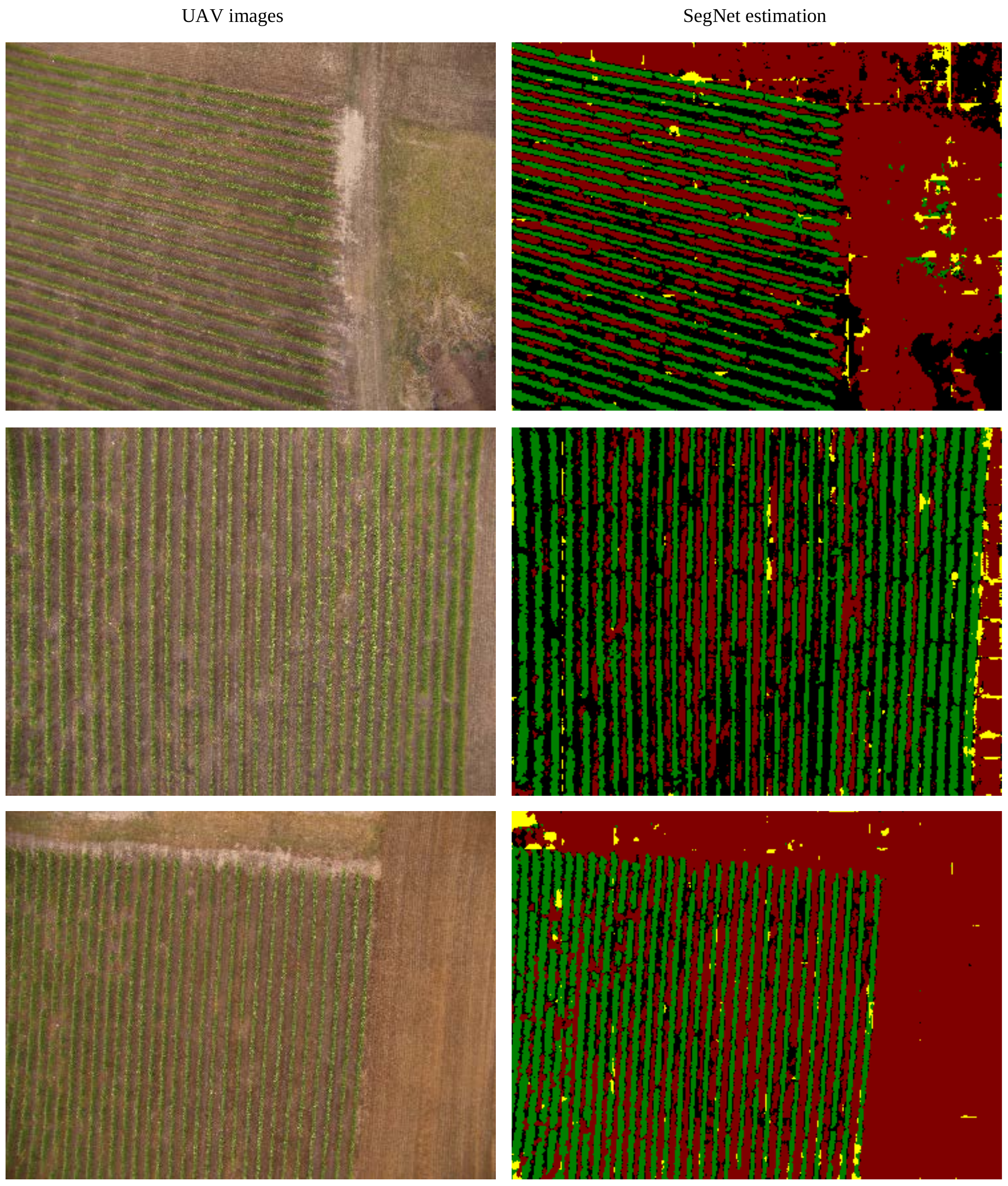}}
	\caption{\label{Qualitatifs_P2} Segmentation qualitative results by the SegNet method for the P2 vineyard. On the left, images from UAV and their right, the segmentation result of these images. The color code of the segmentation is; Black: Shadow, Brown: Ground, Green: Healthy, Yellow: Visible symptom, Orange: Infrared symptom, Red: Symptom intersection.}	
\end{figure}	

\begin{figure}
	\centering{\includegraphics[width=\textwidth]{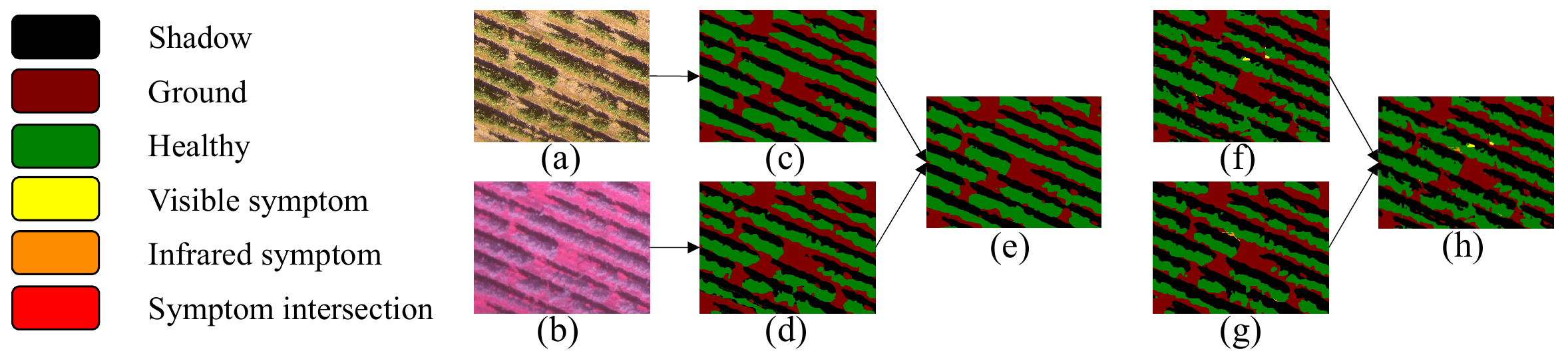}}
	\caption{\label{Healthy_SysExemple} Example of segmentation and fusion of a healthy area. (a): Visible image, (b): Infrared image, (c): Visible ground truth, (d): Infrared ground truth, (e): Fusion ground truth, (f): Visible SegNet estimation, (g): Infrared SegNet estimation, (h): Fusion of segmentation results.}	
\end{figure}	

\begin{figure}
	\centering{\includegraphics[width=\textwidth]{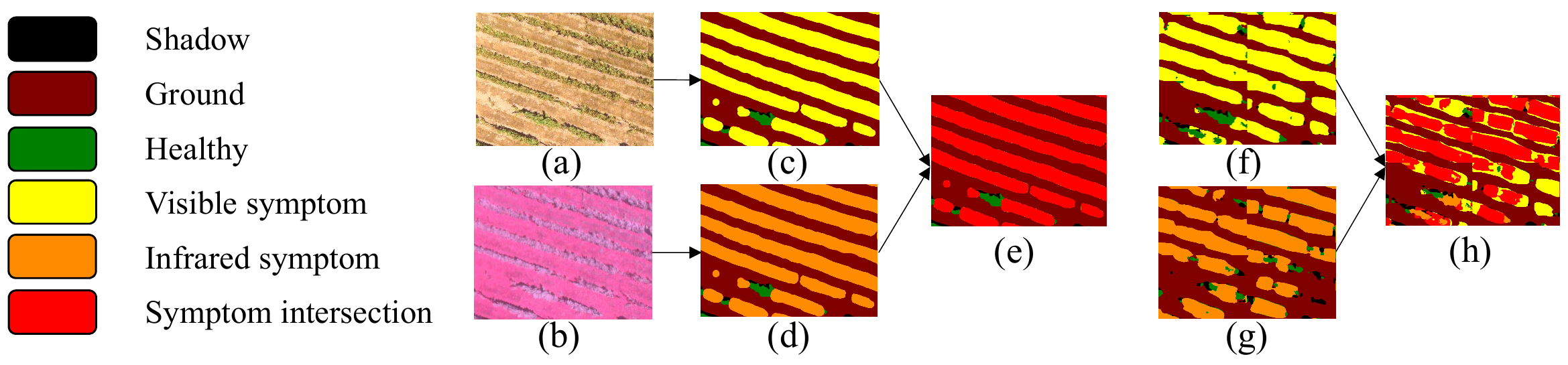}}
	\caption{\label{Diseased_SysExemple} Example of segmentation and fusion of an area contaminated by Mildew. (a): Visible image, (b): Infrared image, (c): Visible ground truth, (d): Infrared ground truth, (e): Fusion ground truth, (f): Visible SegNet estimation, (g): Infrared SegNet estimation, (h): Fusion of segmentation results.}	
\end{figure}

\section{Discussion}
\label{Discussion}
The first question of this study was to determine the ability of multispectral drone imaging to map grapevine Mildew symptoms using machine learning approaches. This led to study imaging modalities in the visible and infrared spectral domains, since several researches have shown the interest of these domains for symptom detection. As our system consists of two cameras for each modality, image alignment was required. We were, therefore, led to develop an algorithm for image registration then using deep learning segmentation to detect the affected surfaces in the vineyard. Another question of this research was, how through the deep learning approach, we can combine the both types of images to delineate symptomatic areas as precisely as possible. Thus, the rate of affected areas can be obtained at the leaf scale or at the vine plant scale. The following sections first discuss the results of the registration and then those of the image segmentation. 

\subsection{Image registration}

In Figure~\ref{ImRegistrationExemple} shows the qualitative result of the registration. On the left and in the middle are the visible and infrared images respectively. Images were taken at a very short time interval (less than one second) and in the same field of view. After the registration of these two images (image on the right) two black areas can be observed on the left and at the bottom of this image. The two areas were captured only by the visible sensor, which explains why there are no equivalents in the infrared image registered at the moment of superposition. There are also areas captured only by the infrared sensor, but due to the fact that the infrared image is registered to the visible image, these areas are not displayed on the result.

An example of how the proposed registration method operates is shown in Figure~\ref{ImRegistrationOptimExemple}. In the image on the left (image registered with the standard method) there is a certain shift between the vinerows present in the visible and infrared images. This is due to a lack of matched points between both images. However, after each iteration, some new correspondence points are detected thanks to the dynamic threshold, and gradually, seven iterations later, a good alignment of the vinerows is achieved as shown in the image on the right.

The quantitative results obtained during the registration experiments are presented in Table~\ref{TabRegistration}. This table reports a statistical study on a dataset of $150$ images, for a comparison between the standard registration method used in~\cite{Sreenu2015, Javadi2015} and the proposed optimized registration method. The values obtained are expressed in "pixels" for the RMSE error measurement, in "seconds" for the runtime and in "times" for the iteration. For each line, the figures in bold show the best results obtained. For standard registration, an average error of $3.29$ pixels was obtained (this result corresponds to the range of results obtained in~\cite{Onyango2017} by the same algorithm), compared to $2.43$ pixels for the optimized method, i.e. an average error reduction of $0.86$ pixels over the entire dataset. This error reduction can be explained by the appearance of new correspondence points, in subsequent iterations, between the two images. These points are used to calculate a new homographic matrix, which corrects the registration slightly, and reduces the RMSE error. However, there are special cases where the optimized method does not reduce the error, in which the result of the optimized method is cancelled and the output of the standard registration is kept. The best score obtained by the standard method yields an error of $1.13$ pixels, whereas with the optimized method the minimum error is $0.87$ pixels. However, both methods can produce larger errors, as we can find errors up to $9.75$ pixels with the standard method and $9.02$ pixels with the optimized method. Note that our disease detection system performs better with an RMSE value $\le 5$; RMSE values between $5$ and $10$ may in some cases reduce the accuracy of disease localization.

The runtime measurement is shown in Table~\ref{TabRegistration}. For the standard registration, an average runtime of $92$ seconds was obtained, versus $139$ seconds with the optimized method. This increase in runtime is due to the additional processing performed by the optimized method ($3.12$ iterations on average). This is not the only reason, however: with the standard method, there is a very significant difference ($80$ seconds) between the minimum and the maximum values. The difference is due to the number of interest points detected in an image by the AKAZE algorithm: the larger the number of interest points detected, the longer the processing time. In other words, there is a direct relation between these two parameters.

The proposed algorithm based on the AKAZE detector was tested by other feature extraction algorithms such as SIFT, SURF, ORB and KAZE. The results obtained by SIFT showed a slightly higher error than those obtained by AKAZE, and the runtime was between 2 to 10 times longer than AKAZE, confirming the study by Tareen et al.~\cite{Saleem2018}. For the other algorithms, several cases of failure were identified. They are mainly due to the lack of correspondence between the two images, which implies a sensitivity of these algorithms to the differences of modality between the two images. Another algorithm developed by Yang et al.~\cite{Dan2018}, which is based on feature extraction through a deep learning architecture was tested, and several problem were observed. Not only, the deep learning method use much of RAM memory, making it very difficult to register high resolution images, but even with a decrease in the resolution of the images, the algorithm cannot find good matches, which leads to the failure of the registration process. This is due to the difference in image modalities.

Some other Area-based registration algorithms were also tested, such as: Normalized Cross-Correlation (NCC)~\cite{Wang2010} and Phase-Correlation (PC)~\cite{Erives2005}. But due mainly to the presence of some deformation on the used dataset images, these algorithms did not manage to correctly register most of the images. Another disadvantage of these algorithms is that they are very time consuming, and were therefore discarded.

\subsection{Image segmentation and fusion}
The qualitative results of the figures~\ref{Qualitatifs_P1} and \ref{Qualitatifs_P2} represent the results on the vineyards of P1 and P2 respectively. For the P1 plot (Figure~\ref{Qualitatifs_P1}), a large part of the untreated plot was contaminated by Mildew. This presence of disease is well detected by segmentation. As can be observed, in most cases, the symptoms are better detected in the visible range (coded by the yellow color) than in the infrared range (coded by the orange color). This hypothesis has been confirmed by the quantitative results in Table~\ref{TabSegResult}. Of course, there are in some cases false detections in both areas. However, when symptoms are detected in both ranges it is likely to be true detection.

For the P2 plot (Figure~\ref{Qualitatifs_P2}), the vineyard is healthy, where the SegNet estimation generally match the ground truth. However, in this example of 3 images, some misclassification of symptomatic areas in the visible range can be observed. This generally occurs when there is a gap in the vegetation (a plant missing in a row, for example) and the color of the soil is similar to that of a symptomatic plant (brown or golden). Also, it is usually explained by the edge effects of the sliding window, because the information is not complete in some border (these effects are not always present in the outputs). In other cases, the symptom class comes out when there is a yellow color mix with green, this case is usually comparable to the symptomatic leaves. To counter this problem, it is necessary to check the neighboring images result which cover this area. If the symptom is detected in all or most of the images, therefore, that can be a real symptom, otherwise, it is a false detection. Misclassification of the other classes can also appear, indeed, in these examples, it can be observed some misclassification of the grass which is detected as healthy vine or symptom class (in the 1st example of Figure~\ref{Qualitatifs_P2}). Also, some confusion between the ground and shadow class because the ground low brightness (in the 2nd and 3rd examples of Figure~\ref{Qualitatifs_P2}).

Figures.~\ref{Healthy_SysExemple} and~\ref{Diseased_SysExemple} show an example of SegNet segmentation and the fusion compared with ground truth. In Figure~\ref{Healthy_SysExemple} the area is healthy, so there are no samples of the symptom class. In this case, it can be observed that the fusion is identical to the visible and infrared estimation (idem for the ground truth). Unlike Figure~\ref{Diseased_SysExemple} which is an area almost completely contaminated by the Mildew, as it can be seen, the ground truth is identical in both spectres, apart from the color code which is different. But for the SegNet output, the result for the symptomatic class detection is not identical in both ranges, which lead to merge the symptomatic visible and infrared classes. In addition, the fusion by intersection is generated by AND operator between the two segmentations output.

The quantitative results obtained for the visible, infrared, fusion by intersection and union segmentation experiments are presented in Table~\ref{TabSegResult} and~\ref{TabDetectionResult}, respectively for leaf-level and grapevine-level. They show the results obtained in terms of the Recall, Precision, F1-Score/Dice coefficient and Accuracy measures, which expressed in percentages.

As shown in the "Accuracy" column of Table~\ref{TabSegResult}, the different classes were generally better detected from the visible image (accuracy of $85.13\%$) than from the infrared image (accuracy of $78.72\%$). This difference in result is due to the fact that the visible image provides a better colorimetric description than the infrared image for the different classes studied. The fusion by union gave a result of $90.23\% $, this result is better than the visible, because the method takes the best of visible and infrared information. It can be noticed that symptoms appear at different locations in the visible and infrared spectral ranges. One interesting finding is that detection in both ranges is complementary, since fusion by union increases the detection performance. The fusion by intersection yields a score of $ 82.20\% $, this result less than the visible range, because the method is conditioned by the intersection of the visible and infrared, and in this case, it is the infrared result that has decreased its result. The fusion by intersection give an important information about the position of the commune symptom detection in visible and infrared. This finding can be used to strengthen the robustness of Mildew detection, where detection can be considered reliable if it is in both types of images. Besides, the fusion by union gives an idea about the quantitative detection.

In addition to measuring the affected areas at leaf level, the second type of assessment consists in testing the detection at vine plant level, as this helps to better manage certain operations in the vineyard. The results present in the Table~\ref{TabDetectionResult} gives a better insight into the symptoms detection at the grapevine-level. The results show that the symptoms detection in the fusion by union is much better (the detection is more than $92.81\%$) compared to the fine scale detection (leaf level), followed by the visible range ($ 91.50\% $), then the infrared range ($81.66\%$). As can be seen, the result of infrared and fusion by intersection are less than pixel-wise evaluation (Table~\ref{TabSegResult}). On the other hand, an increase of precision for the cases of the fusion by union and the visible can be observed.

Results obtained by the proposed method are likely to be consistent with several studies in the field of remote sensing~\cite{Du1998, Jiang2019, Marmanis2018, Chai2019, Images2017, Milioto2018} using the SegNet network. Indeed, the overall accuracy range obtained by these studies is between $70\%$ and $90\%$. In addition, it has been observed that when the surface area is large, the detection result is better. Conversely, the smaller the surface area, the more the SegNet network has trouble to correctly detect the diseased area. A possible explanation for this might be, the lose of the resolution information of small areas during the downsampling and upsampling operations in the SegNet network. Another reason why the results are limited is the difficulty of realizing an accurate ground truth for learning and testing the network, but also the difficulty identifying an area in a low resolution ($ 1cm/pixel $) (in the case of images taken by a UAV at high altitude). In other remote sensing applications with large databases studies that have tested and compared several types of deep learning architecture~\cite{Ma2019a, Chai2019} such as FCN, U-Net, DeepLab, PSPNet, etc., obtained best results from the SegNet network for semantic segmentation.

In the proposed system, the fusion by intersection of the two modalities indicates the locations where symptoms were located at the same position in the visible and infrared images. In other words, fusion provides important information about areas where the system confidence is higher for the disease detection. In other hand, it also provides information about areas where the disease has been detected only in one range (visible or infrared). Therefore, even after the establishment of the method for the diseases detection, the fusion by intersection remains more reliable class than the symptomatic classes detected in the visible or infrared range.

\subsection{Runtime system}
Table~\ref{TabRuntime} presents the runtime results of all stages of the disease detection system. For the image registration step, the average runtime value was used to evaluate the overall system, because the image registration runtime is variable from one pair of images to another. For the SegNet segmentation step, the runtime was multiplied by two ($\times~2$) because the process must segment both images (visible and infrared) and the GPU can only handle one process at a time, unlike registration and fusion operations, where the processing is joint for the two images (visible and infrared). Unlike image registration, the runtime of the SegNet on a UAV image is constant, at $140$ seconds. This value is the same for both visible and infrared images. The fusion between the two segmented images takes less than $2$ seconds; this runtime value includes the computation and saving the fusion file. The runtime of the disease detection system varies according to the image registration method chosen. This implies that for better accuracy of the results, an additional average processing of $47$ seconds per image is necessary.\\

\section{Conclusion}
\label{Conclusion}
In this study, a new method based on optimized images registration and deep learning segmentation method has been proposed for detecting vine disease using multimodal UAV images (visible and infrared ranges). The method consists in three steps. The first one is the images alignment, where an iterative algorithm based on an interest points detector has been developed. The second step is the segmentation of visible and infrared images based on the SegNet architecture to identify four classes: shadow, ground, healthy and symptomatic vine. Lastly, the third step consists in generating a disease map by fusion of the segmentations obtained from the visible and infrared images.
This study showed that the proposed method enables the detection of vine symptoms using information from images of visible and infrared spectra. It provides a framework for the exploration of earlier detection and mapping of the vine diseases. One of the limitations of this research is the small size of the training sample which has reduced the performance of the deep learning segmentation. As future work, improvements could be made. The segmentation method can be improved by enriching the dataset (diversification of disease samples), and also by testing other deep learning architectures for segmentation. Another possibility is the use of 3D information from the vine canopy, thus reducing false detection and improving the accuracy of image registration.

\section*{Acknowledgment}
This work is part of the VINODRONE project supported by the Region Centre-Val de Loire (France). We gratefully acknowledge Region Centre-Val de Loire for its support.\\

\bibliographystyle{unsrt}  
\bibliography{MyBibFile}


\end{document}